\documentclass[11pt,showpacs,preprintnumbers,amsmath,amssymb,prd,nofootinbib,superscriptaddress]{revtex4-2}

\usepackage{dcolumn}% Align table columns on decimal point
\usepackage{bm}% bold math
\usepackage{ifpdf}
\usepackage{hyperref}
\usepackage{bm}
\usepackage{xcolor,color,graphicx,graphics}
\usepackage[spanish,english]{babel}%, portuguese
\usepackage[latin1]{inputenc}
\usepackage[OT1]{fontenc}
\usepackage{latexsym,amssymb,amsmath,amsfonts}
\usepackage{makeidx}
\usepackage{epsfig,subfigure}
\usepackage{natbib}
\usepackage{epstopdf}
\usepackage{mathrsfs}
\usepackage{hyperref}%\usepackage[colorlinks=true,linkcolor=blue]{hyperref}
\hypersetup{colorlinks=true, linkcolor=blue, citecolor=green}
\usepackage{enumerate}

\everymath{\displaystyle}
\usepackage{graphicx}

\usepackage[T1]{fontenc}
\usepackage{amsmath}
\usepackage{amssymb}
\usepackage{graphicx}
\usepackage{xcolor}

\newcommand{\bea}{\begin{eqnarray}}
\newcommand{\eea}{\end{eqnarray}}

\newcommand{\orcid}[1]{\href{https://orcid.org/#1}{\includegraphics[width=10pt]{orcid}}}

\usepackage{fixmath}

\begin{document}

\title{Finite temperature applications in G\"{o}del space-time}

\author{A. F. Santos} %\orcid{0000-0002-2505-5273}}
\email{alesandroferreira@fisica.ufmt.br}
\affiliation{Instituto de F\'{\i}sica, Universidade Federal de Mato Grosso,\\
78060-900, Cuiab\'{a}, Mato Grosso, Brazil}

\author{Faqir C. Khanna \footnote{Professor Emeritus - Physics Department, Theoretical Physics Institute, University of Alberta\\
Edmonton, Alberta, Canada}}
\email{fkhanna@ualberta.ca; khannaf@uvic.ca}
\affiliation{Department of Physics and Astronomy, University of Victoria,\\
3800 Finnerty Road, Victoria BC V8P 5C2, Canada}

\begin{abstract}

Temperature effects in a scalar field non-minimally coupled to gravity are investigated. The Thermo Field Dynamics formalism is used. This is a topological field theory that allows us to calculate different effects, such as the Stefan-Boltzmann law and the Casimir effect, on an equal footing. These phenomena are calculated assuming the G\"{o}del space-time as a gravitational background. A possible implication of these results at the beginning of the universe is discussed.

%\pacs{41.20.Jb, 03.50.De, 03.50.-z, 11.30.Cp}
%\keywords{Electromagnetic wave propagation; Maxwell equations; Lorentz invariance violation}

\end{abstract}

\maketitle

\section{Introduction}

The universe has a non-zero temperature from its beginnings until today. Then the temperature is an important ingredient in nature and must be considered when studying any phenomenon. These thermal effects are present in all known phenomena of the standard model and general relativity. There are two approaches to introducing temperature into a theory. One of them is the imaginary-time formalism which consists of replacing $t$, with a complex time, $i\tau$ \cite{Matsubara}. Then the temporal information of the system is lost. The other is the real-time formalism. In this case, there are two distinct, but equivalent approaches. (i) Closed time path formalism \cite{Schwinger}, where the degrees of freedom of the system are doubled and the Green's function assumes a matrix structure. (ii) Thermo Field Dynamics (TFD) formalism \cite{Umezawa1, Umezawa2, Kbook, Umezawa22, Khanna1, Khanna2}. This procedure is built from the ideas that consider the statistical average of an operator equal to the vacuum expectation value of this operator. For such an interpretation to be possible, two elements are necessary: the Hilbert space must be duplicated and the Bogoliubov transformation is used. Another characteristic of this formalism is its topological structure that allows studying different phenomena on an equal footing, such as the  Stefan-Boltzmann law and the Casimir effect. In this paper, the TFD formalism is used and three different topologies are investigated.

To investigate finite temperature applications, a scalar field coupled to gravity is considered. Then the Stefan-Boltzmann law and the Casimir effect are calculated. G\"{o}del space-time is taken as the gravitational background to develop such a study. The G\"{o}del universe, proposed by Kurt G\"{o}del, in 1949, is an exact solution of Einstein equations that describes a rotating universe \cite{Godel}. The main feature of this metric is that it displays Closed Timelike Curves (CTCs) that allow an observer traveling along them to go back in time. As a consequence, the causality is violated. Although the G\"{o}del universe does not present expansion, several studies with this solution have been developed \cite{Reboucas, Fonseca, Jhonny, Paulo, Paulo2, Santos}. Here, corrections due to this space-time are calculated for the Casimir effect at zero and at finite temperature.

The Casimir effect is a quantum phenomenon proposed by Hendrik Casimir in 1948 \cite{Casimir}. This effect consists of the interaction between two parallel conducting plates placed in the vacuum of a quantum field. This implies modifications in the quantum vacuum and, as a result, the plates are attracted toward each other. The experimental measurement took place ten years after the theoretical prediction \cite{Sparnaay}, and the accuracy has increased over the years  \cite{Lamoreaux, Mohideen}. In the first study, these ideas were considered for the electromagnetic field. However, any quantum field exhibits this effect. Although the Casimir force for a scalar field in the G\"{o}del universe has been calculated \cite{Kho}, in this paper such a study is developed following a different context. Here, the topological structure of the TFD formalism is considered. Then, different topologies, which lead to different compactifications, allow us to obtain the Casimir effect in G\"{o}del space-time at zero and finite temperature. 

This paper is organized as follows. In section II, a brief introduction to the TFD formalism is presented. In section III, the model that describes a massless scalar field non-minimally coupled to gravity is introduced. The energy-momentum tensor associated with the theory is calculated. It is then thermalized to investigate some finite temperature applications. In section IV, the G\"{o}del universe is considered. Using this space-time as a gravitational background, the Stefan-Boltzmann law and the Casimir effect at zero and finite temperature for the scalar field coupled to gravity are calculated. In section V, some concluding remarks are made.

\section{TFD formalism}

TFD is a thermal quantum field theory that takes the statistical average of an arbitrary operator as the expectation value in a thermal vacuum. Such a construction requires two ingredients: the doubling of the Hilbert space and the Bogoliubov transformation. The doubled Hilbert space, i.e. ${\cal S}_T={\cal S}\otimes \tilde{\cal S}$, consist of the original Hilbert space ${\cal S}$ and the dual (tilde) Hilbert space $\tilde{\cal S}$. The Bogoliubov transformation consists of a rotation between tilde and non-tilde operators that introduce the temperature effects. Considering an arbitrary operator ${\cal O}(k)$, this transformation is presented as
\bea
\left( \begin{array}{cc} {\cal O}(k, \alpha)  \\\eta\, \tilde {\cal O}^\dagger(k,\alpha) \end{array} \right)={\cal B}(\alpha)\left( \begin{array}{cc} {\cal O}(k)  \\ \eta\,\tilde {\cal O}^\dagger(k) \end{array} \right),
\eea
where $\eta = -1(+1)$ for bosons (fermions) and  ${\cal B}(\alpha)$ is defined as
\bea
{\cal B}(\alpha)=\left( \begin{array}{cc} u(\alpha) & -v(\alpha) \\
\eta v(\alpha) & u(\alpha) \end{array} \right),
\eea
with $u(\alpha)$ and $v(\alpha)$ being functions given as
\bea
v^2(\alpha)=(e^{\alpha\omega_k}-1)^{-1}, \quad\quad u^2(\alpha)=1+v^2(\alpha).\label{phdef}
\eea
The $\alpha$ is defined as the compactification parameter that can be associated with any physical quantity. Its definition is $\alpha=(\alpha_0,\alpha_1,\cdots\alpha_{D-1})$, where $D$ is the space-time dimension. To introduce the effects of temperature, it is considered as $\alpha_0\equiv\beta$ and $\alpha_1,\cdots\alpha_{D-1}=0$. This parameter can be introduced into the propagator of any quantum field theory. As an example, the scalar field propagator is considered. Then
\bea
G_0^{(AB)}(x-x';\alpha)&=&i\int \frac{d^4k}{(2\pi)^4}e^{-ik(x-x')}G_0^{(AB)}(k;\alpha),
\eea
where $A, B=1, 2$ define the doubled notation and
\bea
G_0^{(AB)}(k;\alpha)={\cal B}^{-1}(\alpha)G_0^{(AB)}(k){\cal B}(\alpha),
\eea
with $G_0(k)$ being the usual massless scalar field propagator. The physical component, which is given by the non-tilde variables ($A=B=1$), is
\bea
G_0^{(11)}(k;\alpha)=G_0(k)+\eta\, v^2(k;\alpha)[G^*_0(k)-G_0(k)],
\eea
where $v^2(k;\alpha)$ is the generalized Bogoliubov transformation. Its definition is given as \cite{GBT}
\bea
v^2(k;\alpha)=\sum_{s=1}^d\sum_{\lbrace\sigma_s\rbrace}2^{s-1}\sum_{l_{\sigma_1},...,l_{\sigma_s}=1}^\infty(-\eta)^{s+\sum_{r=1}^sl_{\sigma_r}}\,\exp\left[{-\sum_{j=1}^s\alpha_{\sigma_j} l_{\sigma_j} k^{\sigma_j}}\right],\label{BT}
\eea
with $d$ being the number of compactified dimensions and $\lbrace\sigma_s\rbrace$ denotes the set of all combinations with $s$ elements.

Furthermore, it is important to emphasize that TFD is a field theory on the topology $\Gamma_D^d=(\mathbb{S}^1)^d\times \mathbb{R}^{D-d}$. Here $1\leq d \leq D$, where $D$ is the dimension of the manifold and $d$ is the number of compactified dimensions. In this formalism any set of dimensions of the manifold $\mathbb{R}^{D}$ can be compactified, where the circumference of the $nth$ $\mathbb{S}^1$ is specified by $\alpha_n$. In this paper, three different topologies will be considered: (i) the topology $\Gamma_4^1=\mathbb{S}^1\times\mathbb{R}^{3}$, where the time-axis is compactified into a circumference of length $\beta$. (ii) The topology $\Gamma_4^1$ with the compactification along the coordinate $z$. (iii) The topology $\Gamma_4^2=\mathbb{S}^1\times\mathbb{S}^1\times\mathbb{R}^{2}$ with a double compactification, with one being the time and one along the coordinate $z$.

\section{The model: scalar field non-minimally coupled to gravity}

The Lagrangian describing a massless scalar field non-minimally coupled to gravity is given as
\bea
{\cal L}=\frac{1}{2}\left(g^{\mu\nu}\partial_\mu\phi(x)\partial_\nu\phi(x)-\xi R\phi(x)^2\right),
\eea
where $\phi(x)$ is the massless scalar field and $\xi$ is the coupling parameter for the scalar curvature, $R$.

In the next section, some finite temperature applications for this theory are investigated. Then the main quantity that must be calculated is the energy-momentum tensor. Its definition, 
\bea
T_{\mu\nu}=-\frac{2}{\sqrt{-g}}\frac{\delta {\cal L}}{\delta g^{\mu\nu}},
\eea
leads to
\bea
T_{\mu\nu}(x)&=&\frac{1}{2}g_{\mu\nu}\partial^\rho\phi(x)\partial_\rho\phi(x)-\partial_\mu\phi(x)\partial_\nu\phi(x)+\xi\left(R_{\mu\nu}-\frac{1}{2}g_{\mu\nu}R+g_{\mu\nu}\Box-\partial_\mu\partial_\nu\right)\phi(x)^2,
\eea
where $R_{\mu\nu}$ is the Ricci tensor and $\Box=g^{\mu\nu}\partial_\mu\partial_\nu$ is the d'Alembert operator. In this quantity there are products of two operators at the same space-time point, this leads to divergence. In order to avoid such problems, the energy-momentum tensor is written at different points in space-time, i.e.
\bea
T_{\mu\nu}(x)&=&\lim_{x'\rightarrow x}\tau\Biggl[\frac{1}{2}g_{\mu\nu}\partial^\rho\phi(x)\partial_\rho\phi(x')-\partial_\mu\phi(x)\partial_\nu\phi(x')\nonumber\\
&+&\xi\left(R_{\mu\nu}-\frac{1}{2}g_{\mu\nu}R+g_{\mu\nu}\Box-\partial_\mu\partial_\nu\right)\phi(x)\phi(x')\Biggl],
\eea
with $\tau$ being the time ordering operator.

Using the canonical quantization for the scalar field
\bea
[\phi(x),\partial'^\mu\phi(x')]=in_0^\mu\delta({\vec{x}-\vec{x'}}),
\eea
and the relation
\bea
\partial^\rho\theta(x_0-x_0')=n_0^\rho\,\delta(x_0-x_0'),
\eea
where $n_0^\mu=(1,0,0,0)$ is a time-like vector and $\theta(x_0-x_0')$ is the step function, the energy-momentum tensor reads
\bea
T_{\mu\nu}(x)&=&\lim_{x'\rightarrow x}\Bigl\{\Delta_{\mu\nu}\tau\left[\phi(x)\phi(x')\right]-\Sigma_{\mu\nu}\delta(x-x')\Bigl\},
\eea
with
\bea
\Delta_{\mu\nu}&=&\frac{1}{2}g_{\mu\nu}\partial^\rho\partial_\rho-\partial_\mu\partial_\nu+\xi\left(R_{\mu\nu}-\frac{1}{2}g_{\mu\nu}R+g_{\mu\nu}\Box-\partial_\mu\partial_\nu\right),\label{Delta}\\
\Sigma_{\mu\nu}&=&-\frac{i}{2}g_{\mu\nu}n_0^\rho\,n_{0\rho}+in_{0\mu}n_{0\nu}.
\eea

The vacuum expectation value of the energy-momentum tensor is
\bea
\left\langle T_{\mu\nu}(x)\right\rangle=\lim_{x'\rightarrow x}\Bigl\{\Delta_{\mu\nu}\left\langle 0\left|\tau[\phi(x)\phi(x')]\right| 0 \right\rangle-\Sigma_{\mu\nu}\delta(x-x')\left\langle 0| 0 \right\rangle\Bigl\}.\label{VEV}
\eea
Considering the massless scalar field propagator $G_0(x-x')$, which is defined as
\bea
\left\langle 0\left|\tau[\phi(x)\phi(x')]\right| 0 \right\rangle=iG_0(x-x'),
\eea
Eq. (\ref{VEV}) becomes
\bea
\left\langle T_{\mu\nu}(x)\right\rangle=\lim_{x'\rightarrow x}\Bigl\{i\Delta_{\mu\nu}G_0(x-x')-\Sigma_{\mu\nu}\delta(x-x')\Bigl\}.\label{VEV1}
\eea

Applying the TFD formalism, the $\alpha$ parameter is introduced. Then the vacuum expectation value of the energy-momentum tensor is given as
\bea
\left\langle T_{\mu\nu}^{(AB)}(x;\alpha)\right\rangle=\lim_{x'\rightarrow x}\Bigl\{i\Delta_{\mu\nu}G_0^{(AB)}(x-x';\alpha)-\Sigma_{\mu\nu}\delta(x-x')\delta^{(AB)}\Bigl\}.
\eea

In order to study some application, a physical (renormalized) energy-momentum tensor is needed. Then the following prescription is used
\bea
{\cal T}_{\mu\nu}(x;\alpha)=\left\langle T_{\mu\nu}^{(AB)}(x;\alpha)\right\rangle-\left\langle T_{\mu\nu}^{(AB)}(x)\right\rangle.
\eea
This leads to
\bea
{\cal T}_{\mu\nu}(x;\alpha)=\lim_{x'\rightarrow x}\Bigl\{i\Delta_{\mu\nu}\overline{G}_0^{(AB)}(x-x';\alpha)\Bigl\},
\eea
with
\bea
\overline{G}_0^{(AB)}(x-x';\alpha)=G_0^{(AB)}(x-x';\alpha)-G_0^{(AB)}(x-x').\label{Green}
\eea

In the next section, these results are used to investigate some finite temperature applications in the G\"{o}del universe.

\section{G\"{o}del space-time and applications}

Here the main characteristics of the G\"{o}del universe are presented. The G\"{o}del metric is given as
\bea
ds^2=a^2\left[dt^2-dx^2+\frac{1}{2}e^{2x}dy^2-dz^2+2e^xdt dy\right],
\eea
where $a$ is a positive constant. The relevant quantities for the following calculations are: (i) non-zero Ricci tensor components:
\bea
R_{00}=1,\quad\quad\quad R_{02}=R_{20}=e^x, \quad\quad\quad R_{22}=e^{2x},
\eea
and (ii) the Ricci scalar:
\bea
R=\frac{1}{a^2}.
\eea
Using these elements, the components with $\mu=\nu=0$ and $\mu=\nu=3$ on Eq. (\ref{Delta}) are given as
\bea
\Delta_{00}&=&-\frac{1}{2}\left(3\partial_0\partial'_0-4e^{-x}\partial_0\partial'_2+\partial_1\partial'_1+2e^{-2x}\partial_2\partial'_2+\partial_3\partial'_3\right)\nonumber\\
&+&\xi\left(\frac{1}{2}-\left(2\partial_0\partial'_0-4e^{-x}\partial_0\partial'_2+\partial_1\partial'_1+2e^{-2x}\partial_2\partial'_2+\partial_3\partial'_3\right)\right)\label{00}
\eea
and
\bea
\Delta_{33}&=&\frac{1}{2}\left(\partial_0\partial'_0-4e^{-x}\partial_0\partial'_2+\partial_1\partial'_1+2e^{-2x}\partial_2\partial'_2-\partial_3\partial'_3\right)\nonumber\\
&+&\xi\left(\frac{1}{2}+\partial_0\partial'_0-4e^{-x}\partial_0\partial'_2+\partial_1\partial'_1+2e^{-2x}\partial_2\partial'_2\right).\label{33}
\eea

The scalar field propagator in this gravitational background is given by
\bea
G_0(x-x')=-\frac{i}{(2\pi)^2}\frac{1}{(x-x')^2},
\eea
where
\bea
(x-x')^2&=&a^2(t-t')^2+2a^2e^x(t-t')(y-y')-a^2(x-x')^2\nonumber\\
&+&\frac{1}{2}a^2e^{2x}(y-y')^2-a^2(z-z')^2.
\eea

Now using these ingredients and different topologies, some applications at finite temperature are calculated.

\subsection{Stefan-Boltzmann law in G\"{o}del space-time}

In order to obtain the energy density in G\"{o}del universe the topology $\Gamma_4^1=\mathbb{S}^1\times\mathbb{R}^{3}$, with $\alpha=(\beta,0,0,0)$ is considered. In this case the generalized Bogoliubov transformation is given as
\bea
v^2(\beta)=\sum_{l_0=1}^{\infty}e^{-\beta k^0l_0},\label{BT1}
\eea
and the Green function is
\bea
\overline{G}_0(x-x';\beta)=2\sum_{l_0=1}^{\infty}G_0(x-x'-i\beta l_0n_0),\label{GF1}
\eea
where $\overline{G}_0(x-x';\beta)\equiv \overline{G}_0^{(11)}(x-x';\beta)$ represents the physical component and $n_0=(1,0,0,0)$. Then the energy-momentum tensor becomes
\bea
{\cal T}^{(11)}_{\mu\nu}(x;\beta)&=&2i\lim_{x'\rightarrow x}\Bigl\{\Delta_{\mu\nu}\sum_{l_0=1}^{\infty}G_0(x-x'-i\beta l_0n_0)\Bigl\}.
\eea
Taking $\mu=\nu=0$ and using Eq. (\ref{00}) in the last equation, we get
\bea
{\cal T}^{(11)}_{00}(T)=\frac{\pi^2}{30a^2}T^4+\frac{\xi}{a^2}\left(\frac{\pi^2}{30}-\frac{1}{96T^2}\right)T^4.
\eea
This is the Stefan-Boltzmann law associated with the scalar field coupled to gravity in the G\"{o}del space-time. The G\"{o}del parameter $a^2$ changes the usual result. If the coupling constant $\xi$ is very small, the standard result for the massless scalar field is recovered if $a^2=1$. It is interesting to observe two temperature limits: at very high temperature $E\propto T^4$, while at very low temperature $E\propto T^2$.

\subsection{Casimir effect in G\"{o}del space-time}

In this subsection the Casimir effect for the massless scalar field coupled to gravity in G\"{o}del space-time is calculated in two different topologies. This leads to obtaining the Casimir effect at zero and finite temperature.

\subsubsection{Zero temperature}

To calculate the Casimir effect in G\"{o}del space-time at zero temperature the topology $\Gamma_4^1=\mathbb{S}^1\times\mathbb{R}^{3}$ with $\alpha=(0,0,0,i2d)$ is considered. Here $2d$ is the length of the circumference $\mathbb{S}^1$. Then the Bogoliubov transformation is given as
\bea
v^2(d)=\sum_{l_3=1}^{\infty}e^{-i2d k^3l_3}\label{BT2}
\eea
and the Green function is written as
\bea
\overline{G}_0(x-x';d)=2\sum_{l_3=1}^{\infty}G_0(x-x'-2d l_3n_3)\label{GF2}
\eea
with $n_3=(0,0,0,1)$. Using these results, the energy-momentum tensor becomes
\bea
{\cal T}^{(11)}_{\mu\nu}(x;d)&=&2i\lim_{x'\rightarrow x}\Bigl\{\Delta_{\mu\nu}\sum_{l_3=1}^{\infty}G_0(x-x'-2d l_3n_3)\Bigl\}.
\eea
Choosing $\mu=\nu=0$ the Casimir energy in G\"{o}del space-time reads
\bea
{\cal T}^{(11)}_{00}(d)=-\frac{\pi^2}{1440a^2d^4}-\xi\left(\frac{\pi^2}{1440a^2d^4}+\frac{1}{96a^2d^2}\right).
\eea
And for $\mu=\nu=3$ the Casimir pressure is given as
\bea
{\cal T}^{(11)}_{33}(d)=-\frac{\pi^2}{480a^2d^4}-\xi\left(\frac{\pi^2}{480a^2d^4}+\frac{1}{96a^2d^2}\right).
\eea
Therefore the G\"{o}del space-time allows for the Casimir effect at zero temperature. The gravitational background, described by the G\"{o}del metric, changes the usual Casimir effect for the massless scalar field coupled with gravity. The modifications due to the gravitational background are governed by the parameter $a^2$. Three cases can be analyzed. (i) $a^2=1$ leads to the standard result; (ii) $a^2$ very small implies that the Casimir effect increases; and (iii) for $a^2$ very large the Casimir effect in the G\"{o}del universe decreases compared to the standard result.

\subsubsection{Finite temperature}

To investigate thermal corrections for the Casimir effect in G\"{o}del space-time two compactifications are considered, i.e., one along the time axis and the other along the $z$-coordinate. The topological structure that represents this effect is given by $\Gamma_4^2=\mathbb{S}^1\times\mathbb{S}^1\times\mathbb{R}^{2}$. In this case the $\alpha$ parameter is chosen as $\alpha=(\beta,0,0,i2d)$. For this choice, the Bogoliubov transformation is given as
\bea
v^2(\beta,d)=\sum_{l_0=1}^\infty e^{-\beta k^0l_0}+\sum_{l_3=1}^\infty e^{-i2dk^3l_3}+2\sum_{l_0,l_3=1}^\infty e^{-\beta k^0l_0-i2dk^3l_3},\label{BT3}
\eea
where $dk^3$ is the product between the distance $d$ and the $z$-component of the momentum $k^3$.
It is important to note that the first term leads to the Stefan-Boltzmann law and the second term is related to the Casimir effect at zero temperature. The third term is associated with the Casimir effect with non-zero temperature. In this case, two compactifications happen together. The Green function that describes the third term is
\bea
\overline{G}_0(x-x';\beta,b)&=&4\sum_{l_0,l_3=1}^\infty G_0\left(x-x'-i\beta l_0n_0-2dl_3n_3\right).\label{GF3}
\eea

Using this result, the energy-momentum tensor becomes
\bea
{\cal T}^{(11)}_{\mu\nu}(x;\beta,d)&=&4i\lim_{x'\rightarrow x}\Bigl\{\Delta_{\mu\nu}\sum_{l_0,l_3=1}^\infty G_0\left(x-x'-i\beta l_0n_0-2dl_3n_3\right)\Bigl\}.
\eea
After some calculations, the Casimir energy for the scalar field coupled to gravity at finite temperature in G\"{o}del space-time is given as
\bea
{\cal T}^{(11)}_{00}(\beta,d)&=&-\frac{2}{\pi^2 a^2}\sum_{l_0,l_3=1}^\infty\Biggl\{\frac{(2dl_3)^2-3(\beta l_0)^2}{[(2dl_3)^2+(\beta l_0)^2]^3}\nonumber\\
&-&\frac{\xi}{4}\frac{[-(4dl_3)^2(1+(dl_3)^2)+(2\beta l_0)^2(3-2(dl_3)^2)-(\beta l_0)^4]}{[(2dl_3)^2+(\beta l_0)^2]^3}\Biggl\}.
\eea
Similarly, the Casimir pressure at finite temperature in G\"{o}del universe reads
\bea
{\cal T}^{(11)}_{33}(\beta,d)&=&-\frac{2}{\pi^2 a^2}\sum_{l_0,l_3=1}^\infty\Biggl\{\frac{3(2dl_3)^2-(\beta l_0)^2}{[(2dl_3)^2+(\beta l_0)^2]^3}\nonumber\\
&-&\frac{\xi}{4}\frac{[-(4dl_3)^2(3+(dl_3)^2)+(2\beta l_0)^2(1-2(dl_3)^2)-(\beta l_0)^4]}{[(2dl_3)^2+(\beta l_0)^2]^3}\Biggl\}.
\eea
These expressions show that the Casimir energy and Casimir pressure are affected by the effects of temperature in addition to modifications due to the G\"{o}del parameter.

\section{Conclusions}

A scalar field coupled to gravity in the TFD formalism is considered. TFD is an approach that allows introducing temperature effects using a topological structure. Different topologies lead to different phenomena, although all effects are treated on an equal footing. Here, three topologies are chosen which imply: (i) the time is compactified into a circumference of length $\beta$, (ii) the $z$-coordinate is compactified, and (iii) both compactifications are taken together. As a result, three different effects associated with the scalar field non-minimally coupled to gravity are calculated: the Stefan-Boltzmann law and the Casimir effect at zero and non-zero temperature. In such a study, G\"{o}del space-time is considered. This cosmological model is an exact solution of general relativity. Its main feature is the possibility of CTCs that leads to a violation of causality. Our results show that the G\"{o}del universe allows for a Stefan-Boltzmann law and the Casimir effect. However, these phenomena depend on the G\"{o}del parameter $a^2$. Although the G\"{o}del universe does not describe expansion, then it is not a realistic model of our actual universe, there is an important discussion that considers that the universe can experience a G\"{o}del phase for a small period of time during the inflation era. This characterizes the de Sitter-G\"{o}del-de Sitter phase transition, i.e., there is a phase transition from de Sitter to G\"{o}del space-time and then back to de Sitter space-time \cite{Kho, Kho2}. Therefore, the study developed here is important since the Casimir effect in G\"{o}del space-time is a quantum effect and may be relevant at the beginning of the universe. Furthermore, the study developed here can be generalized considering the G\"{o}del-type space-time. Using cylindrical coordinates, the causality problem is examined in more detail. Causal and non-causal regions are allowed. Then it is interesting to investigate how the parameters that determine such regions affect the Casimir effect at zero and finite temperature. This analysis is under investigation in work-in-progress.

\section*{Acknowledgments}

This work by A. F. S. is supported by National Council for Scientific and Technological Develo\-pment - CNPq projects 430194/2018-8 and 313400/2020-2.

\end{document}